\documentclass[11pt, letterpaper]{article}
\usepackage[letterpaper,left=1in,right=1in,top=1in,bottom=1in]{geometry}
\usepackage{graphicx}
\usepackage{amsmath}
\usepackage{amssymb}
\usepackage{setspace}
\usepackage{epstopdf}
\usepackage{color}
\usepackage{booktabs}
\usepackage{multirow}
\usepackage{cite}
\usepackage{siunitx,etoolbox} 
\usepackage{arydshln}
\usepackage{subfigure}
\newtheorem{rmk}{Remark}
\usepackage{algorithmic}
\usepackage{algorithm}

\title{\Large \bf Self-tuning moving horizon estimation of nonlinear systems via physics-informed machine learning Koopman modeling}

\author{
\centerline{\normalsize Mingxue Yan$^{a,b,d}$, Minghao Han$^{a}$, Adrian Wing-Keung Law$^{c,}$\thanks{Corresponding author: A. W.-K. Law. Tel: (+65) 6516 2273. Email: cewklaw@nus.edu.sg.}$~$, Xunyuan Yin$^{a,d,}$\thanks{Corresponding author: X. Yin. Tel: (+65) 6316 8746. Email: xunyuan.yin@ntu.edu.sg.}
}
\vspace{5mm}\\
\centerline{\small $^{a}$ Nanyang Environment and Water Research Institute (NEWRI),}\\
\centerline{\small Nanyang Technological University, 1 CleanTech Loop, 637141, Singapore}\\
\centerline{\small $^{b}$ Interdisciplinary Graduate Programme, Nanyang Technological University,}\\
\centerline{\small 61 Nanyang Drive, 637460, Singapore}\\
\centerline{\small $^{c}$ Department of Civil and Environmental Engineering, National University of Singapore,}\\
\centerline{\small 1 Engineering Drive 2, 117576, Singapore}\\
\centerline{\small $^{d}$ School of Chemistry, Chemical Engineering and Biotechnology, Nanyang Technological University,}\\
\centerline{\small 62 Nanyang Drive, 637459, Singapore}\\}
\allowdisplaybreaks

\date{}

\begin{document}
\maketitle
\setstretch{1.5}

\begin{abstract}
In this paper, we propose a physics-informed learning-based Koopman modeling approach and present a Koopman-based self-tuning moving horizon estimation design for a class of nonlinear systems. Specifically, we train Koopman operators and two neural networks - the state lifting network and the noise characterization network - using both data and available physical information. The first network accounts for the nonlinear lifting functions for the Koopman model, while the second network characterizes the system noise distributions. Accordingly, a stochastic linear Koopman model is established in the lifted space to forecast the dynamic behaviors of the nonlinear system. Based on the Koopman model, a self-tuning linear moving horizon estimation (MHE) scheme is developed. The weighting matrices of the MHE design are updated using the pre-trained noise characterization network at each sampling instant. The proposed estimation scheme is computationally efficient, since only convex optimization needs to be solved during online implementation, and updating the weighting matrices of the MHE scheme does not require re-training the neural networks. We verify the effectiveness and evaluate the performance of the proposed method via the application to a simulated chemical process.


\end{abstract}

\section{Introduction}

Koopman operator-based data-driven modeling and model predictive control (MPC) has become an emerging integrated modeling and optimal control framework for nonlinear systems, particularly in cases where high-fidelity mechanistic models are absent or accurate model parameters are unavailable. According to Koopman theory, for any nonlinear system, it is possible to find a higher-dimensional space where a Koopman operator can be identified to describe the dynamic evolution of the original nonlinear system in a linear manner~\cite{koopman1931hamiltonian,kp1}. Within a Koopman operator framework, linear MPC strategies can be proposed and applied for nonlinear systems, and complex and time-consuming nonlinear optimization associated with conventional nonlinear MPC methods can be bypassed. 

A variety of Koopman-based MPC algorithms have been developed. In~\cite{kp2,folkestad2021koopman,arbabi2018data}, Koopman operators were used to approximate the dynamic behaviors of deterministic nonlinear systems, leading to the development of deterministic linear convex MPC based on the obtained Koopman model. 
In~\cite{Li2024CCE}, a nominal MPC was developed based on an input-augmented Koopman model which takes into account the nonlinear dependence of the control input.
To account for system disturbances and the inherent plant-model mismatch, stochastic Koopman models were established to enable robust Koopman MPC implementation~\cite{zhang2022robust,han2023robust,desko}.
In~\cite{zhang2022robust}, tube-based MPC was proposed to address nonlinear discrete systems with potential exogenous disturbances and noise, which offers greater tolerance for modeling approximation errors. 
In~\cite{han2023robust}, stochastic systems were modeled using a deep stochastic Koopman operator. In~\cite{desko}, Koopman modeling for stochastic systems with time delays was conducted by using long short-term memory (LSTM) to account for the dependence of the current states on historical states.
In~\cite{narasingam2019koopman}, the Koopman model was integrated with Lyapunov-based MPC for the stabilization of nonlinear systems. In~\cite{son2021application, muske2002disturbance}, the offset-free control framework was integrated with Koopman-based MPC, which compensates for both model mismatch and process disturbances. In~\cite{son2022development}, offset-free Koopman Lyapunov-based MPC was proposed to address the modeling mismatch and ensure the closed-loop stability of the control system.
Additionally, in~\cite{kempc}, a Koopman-based economic MPC scheme was proposed to achieve better economic operational performance. While the Koopman-MPC framework has been well studied, a critical remaining issue is the requirement for full-state measurements for online decision-making in most Koopman-based MPC methods. In real applications, measuring some key variables online using hardware sensors can be challenging~\cite{yin2019subsystem,li2023partition}. This necessitates state estimation, which provides real-time estimates of the necessary quality variables based on limited output measurements~\cite{yin2018state,duan2020nonlinear, yin2018forming, tang2023data, minmaxmhe, tang2021nonlinear,alhajeri2021machine}. In this work, we aim to leverage the Koopman operator concept to develop a computationally efficient state estimation method for nonlinear systems, which can be integrated with Koopman MPC for efficient monitoring and optimal operation of general nonlinear systems.

Various data-driven Koopman modeling approaches have been proposed to build models suitable for implementing linear estimation and control algorithms. Extended dynamic mode decomposition (EDMD)~\cite{edmd} is one of the most representative methods. This type of method lifts the system state to a high dimensional space through manually selected lifting functions such as basis functions and reproducing kernels. Then, a least-squares problem is solved to build Koopman operators, and the resulting linear state-space model in the higher-dimensional space can be leveraged for linear estimation and control~\cite{edmd, son2022hybrid}. However, selecting appropriate lifting functions may require extensive experience and domain knowledge about the underlying dynamics of each specific process~\cite{dmdkernel,edmd}. It is also challenging to find suitable lifting functions through manual selection when dealing with large-scale systems~\cite{yin2023data, edmd}. 

In our recent work~\cite{yin2023data}, we made an initial attempt on Koopman-based constrained state estimation for nonlinear systems using EDMD-based modeling. While promising results were achieved, scaling this approach to systems with numerous state variables remains constrained by the difficulty of manually selecting extensive lifting functions. To address this challenge, learning-based approaches to Koopman modeling have been proposed, where neural networks embed the system state into a lifted space. This type of method automates the selection of lifting functions by training Koopman operators on offline data~\cite{learningKoopman}, which holds the promise to facilitate broader adoption of Koopman modeling for nonlinear system estimation and control.
However, the effectiveness of learning-based Koopman models heavily relies on the quantity and quality of available data. Achieving accurate approximation and generalization demands diverse and sufficiently large training datasets. Neural networks can struggle with small-scale datasets, leading to overfitting or convergence issues. Moreover, the performance of neural network-based lifting functions and resulting Koopman models can be significantly impacted by noise and outliers in the data~\cite{liano1996robust}.

In machine learning, models that integrate the physical knowledge and the data information have the potential to address challenges related to data scarcity and poor data quality. One possible approach is to build a hybrid model that integrates the first-principles model and a trainable neural network~\cite{lee2020development,psichogios1992hybrid, bangi2020deep} explicitly. Neural networks are used to directly compensate for discrepancies between the first-principles model and the physical plant~\cite{lee2020development} or to estimate certain unknown process parameters of a first-principles model~\cite{psichogios1992hybrid, bangi2020deep}. Another promising approach is to incorporate available valuable physical knowledge into neural network training, which is known as physics-informed neural networks (PINNs)~\cite{piml}. PINN-based methods have been widely used for approximating partial differential equations (PDEs) by integrating data and mathematical models to enforce physical laws. PINN has also been considered in addressing control-oriented machine learning modeling for nonlinear ordinary differential equations (ODEs) processes. The use of a PINN-based dynamic model can lead to improved control performance compared with using ODEs and classical integration 
methods such as fourth-order Runge-Kutta method (RK4)\cite{pinnMPC,pinnMPC2,xiao2023modeling}.
PINNs has the potential to be integrated with the learning-based Koopman modeling framework, which is particularly relevant for scenarios where data may be scarce for purely data-driven, data-intensive deep learning-based Koopman modeling. By incorporating available physical knowledge into the training process, PINNs can help develop more appropriate neural network-based lifting functions.
As discussed previously, purely data-driven models that rely solely on data might fit the training data well. However, the predictions from these models can deviate from real system behaviors due to biases in the collected data, which will lead to poor generalization performance~\cite{ying2019overview}. Incorporating valuable physical information allows neural networks to maintain robustness when conducting machine learning modeling on noisy datasets~\cite{piml}. Moreover, when data availability is limited, the exploration space of a trained neural network is constrained, and the risk of overfitting is increased. Integrating data and physical information in a unified approach is promising for mitigating overfitting tendencies, resulting in more physically consistent and robust predictions. 

Based on the above observations and considerations, in this work, we aim to propose a learning-based Koopman modeling and state estimation approach for general nonlinear systems under practical scenarios where data can be limited and/or noisy. We propose a physics-informed machine learning Koopman modeling approach, which leverages limited data and the available physical knowledge to build a stochastic predictive Koopman model to robustly characterize the dynamics of the considered nonlinear system. 
A learning-enabled MHE scheme with self-tuning weighting matrices is developed based on the Koopman model. In this scheme, the weighting matrices of the Koopman MHE are updated online using a pre-trained noise characterization network to reduce the efforts of manual parameter tuning associated with conventional MHE designs. This estimation scheme solves a convex optimization problem at each sampling instant to online estimate the state of the underlying nonlinear system efficiently. A benchmark reactor-separator process example is introduced to illustrate the proposed approach. 
Our method showcases its capability to build an accurate Koopman model using less data than required for purely data-driven Koopman modeling methods. Moreover, the self-tuning of weighting matrices in the MHE leads to more precise state estimates as compared to conventional designs where weighting matrices remain constant.

\section{Preliminaries and problem formulation}
\label{sec:preliminaries}

\subsection{Notation}
$x_{i:j}$ is a sequence of vector $x$ from time instant $i$ to $j$, $x_{j|k}$ is the information about $x$ for time instant $j$ obtained at time instant $k$. $\left\|x\right\|^2$ represents the Euclidean norm of vector $x$. $\left\|x\right\|^2_P$ is the square of the weighted norm of vector $x$, that is, $\left\|x\right\|^2_P = x^\top P x$. $\text{diag}\left(x\right)$ represents a diagonal matrix of which the $i$th main diagonal element is constituted by the $i$th element of vector $x$. $f^p={[f_i, f_j]}^{\top}$ denotes a new vector consisting of the $i$th and $j$th elements of $f$. $\mathcal{N}(a,b)$ denotes a Gaussian distribution with mean $a$ and variance $b$. $I_{n}$ is an identity matrix of size $n\times n$. $\bf{0}$ denotes a null matrix of appropriate size. 

\subsection{Koopman operator for controlled systems}
 
Consider a general discrete-time nonlinear system as follows:
\begin{equation}
\label{paper1:K:nonliear}
    x_{k+1} = f(x_k,u_k)
\end{equation}
where $k$ represents the sampling instant; $x \in \mathbb{X} \subseteq \mathbb{R}^{n_x} $ denotes the system state; $u \in \mathbb{U} \subseteq \mathbb{R}^{n_u}$ represents the control input. 

Based on the Koopman theory for controlled systems, there exists an infinite-dimensional space $\mathcal{G}$, where the dynamics of (\ref{paper1:K:nonliear}) can be described using a linear model~\cite{kp2}. 
However, from a practical viewpoint, finding the exact infinite-dimensional lifting functions $\Psi$, and Koopman operator for nonlinear system (\ref{paper1:model:nonlinear}) can be infeasible. Instead, constructing a finite-dimensional approximation of the exact Koopman operator can be a practical solution for real-world nonlinear systems. 
Specifically, let us consider an augmented state vector $\mathcal X$ defined as 
$\mathcal X = \big[ x_k^{\top},u_k^{\top} \big]^{\top}$. We can use finite-dimensional lifting functions $\Psi$ and establish an approximated Koopman operator, denoted by $\mathcal{K}$, within the corresponding finite-dimensional state space, such that:
\begin{equation}
\label{paper1:approx:Koopman}
\Psi (\mathcal X_{k+1}) \approx  \mathcal{K}
\Psi (\mathcal X_{k})
\end{equation}

According to \cite{kp2}, the lifting functions (or observables) can be made with the following form:
\begin{equation}
\label{paper1:approx:Psi}
\Psi\left(\left[
 \begin{array}{c}
 x\\
 u\\
 \end{array}
 \right]\right) = \left[
 \begin{array}{c}
 g(x)\\
 u\\
 \end{array}
 \right]
\end{equation}

We note that the approximated Koopman operator $\mathcal{K}$ in (\ref{paper1:approx:Koopman}) can be partitioned into four blocks as follows:
\begin{equation}
\label{paper1:K}
    \mathcal{K} = \left[
    \begin{array}{c;{1.5pt/1.5pt}c}
        A&B \\ \hdashline[2pt/2pt]
        *&*
    \end{array}
\right]
\end{equation}

Since our focus is on establishing a Koopman model for forecasting the future behavior of state $x$ rather than predicting the future behavior of the control input $u$, it is sufficient to build a Koopman model that describes the dynamic behavior of $g(x)$, which contains the future information of stat. Consequently, we only need to reconstruct matrices $A$ and $B$ and build a linear Koopman-based state-space model in the following form:
\begin{equation}
\label{paper1:model:Koopman}
    g(x_{k+1}) = A g(x_k)+Bu_k
\end{equation}
where $A \in \mathbb{R}^{n_g \times n_g}$, $B \in \mathbb{R}^{n_g \times n_u}$ are sub-matrices of the finite Koopman operator $\mathcal{K}$; $g \in \hat{\mathcal{G}} \subseteq \mathbb{R}^{n_g}$ denotes the observables in the finite-dimensional lifted space $\hat{\mathcal{G}}$. Further, based on the observables $g$, we can obtain the reconstructed state $\hat x_k$ by using a reconstruction matrix $C$ following:
\begin{equation}
\label{paper1:model:x}
    \hat x_k=Cg(x_k).
\end{equation}

\subsection{Problem statement}
\label{sec:Problem statement}
Consider a general stochastic discrete-time nonlinear system described by the following state-space model:
\begin{subequations}\label{paper1:model:nonlinear}
\begin{align}
    x_{k+1} &= f(x_k,u_k) + w_k \label{paper1:model:nonlinear:1}\\
    y_k &= h(x_k)+v_k\label{paper1:model:nonlinear:2}
\end{align}
\end{subequations}
where $k$ represents the sampling instant; $x \in \mathbb{X} \subseteq \mathbb{R}^{n_x} $ denotes the system state; $u \in \mathbb{U} \subseteq \mathbb{R}^{n_u}$ is the control input; $y \in \mathbb{Y} \subseteq \mathbb{R}^{n_y}$ is vector of output measurements; $w \in \mathbb{W} = \{\  w\subseteq \mathbb{R}^{n_x} ~\text{s.t.}~ ||w||_2\leq w_{\max}\}$ and $v \in \mathbb{V} = \{\ v\subseteq \mathbb{R}^{n_y} ~\text{s.t.}~||v||_2\leq v_{\max}\}$ represent the system disturbances and measurement noise respectively, with $w_{\text{max}}$ and $v_{\text{max}}$ being the supremum of the norms of $w$ and $v$; $f:\mathbb{X}\times\mathbb{U}\to\mathbb{X}$ is a nonlinear vector function that describes the dynamic behavior of $x$, and is typically established through first-principles modeling; $h:\mathbb{X}\to\mathbb{Y}$ is the output measurement function.

In this work, we treat the problem of state estimation of the stochastic nonlinear system in~(\ref{paper1:K:nonliear}), under practical case scenarios when accurate details of the first-principles process model, denoted as $f$, and/or the values of the associated model parameters are only partially available. To address the nonlinear state estimation problem in a computationally efficient manner, we aim to leverage both the system data, and available physical knowledge about the expression and the parameter values of $f$, to build a stochastic Koopman-based linear dynamic model, which can approximate the dynamic behaviors of~(\ref{paper1:K:nonliear}), formulated as follows:
\begin{subequations}\label{paper1:model:Linear}
\begin{align}
z_{k+1} &= Az_k + Bu_k + \mu_k  \label{paper1:model:Linear:1} \\
\hat{x}_k &= Cz_k  \label{paper1:model:Linear:2} \\
\hat {y}_k &= Dz_k \label{paper1:model:Linear:3}
\end{align}
\end{subequations}
where $z = g(x) \in \hat{\mathcal{G}} \subseteq \mathbb{R}^{n_g}$ represents the lifted state vector in the context of the linear Koopman model; $\mu\subseteq \mathbb{R}^{n_g}$ is the disturbance to the lifted state; $\hat x \subseteq \mathbb{R}^{n_x}$ and $\hat y \subseteq \mathbb{R}^{n_y}$ denote the reconstructed system state and output, respectively; $A \in \mathbb{R}^{n_g \times n_g}$, $B \in \mathbb{R}^{n_g \times n_u}$, $C \in \mathbb{R}^{n_x \times n_g}$, and $D \in \mathbb{R}^{n_y \times n_g}$ are system matrices to be established from data.

Then, based on the stochastic Koopman model, a linear moving horizon estimation scheme will be developed to realize constrained estimation of the full state $x$ of~(\ref{paper1:model:nonlinear}) in a linear manner, despite the nonlinearity of the underlying system dynamics.

\section{Physics-informed learning-based Koopman modeling}
\label{sec:PIMLmodeling}
\subsection{Existing approaches to lifting function selection}
An appropriate selection of the lifting functions is critical for building an accurate Koopman model with good predictive capabilities. In the existing literature, various methods have been proposed for obtaining finite-dimensional approximations of the Koopman operator and determining the corresponding lifting functions. 

Extended dynamic mode decomposition is one of the representative approaches~\cite{edmd}. Within the EDMD framework, the lifting functions are manually selected based on trial-and-error analysis, domain knowledge, and users' experience. Despite its promise in handling high nonlinearity, the necessity for manual selection restricts the application of EDMD-based Koopman modeling to medium- to large-scale nonlinear systems. Koopman modeling based on Kalman-Generalized Sparse Identification of Nonlinear Dynamics (Kalman-GSINDy) proposed in \cite{kp1} creates a rich library containing a comparatively large number of candidate lifting functions, and a Kalman-based algorithm is executed recursively to select the most relevant lifting functions from the rich library to account for the nonlinear mapping. However, the creation of this library containing candidate lifting functions remains a non-trivial task, and there may not exist a generalized library that is applicable to a broad range of nonlinear systems.

The learning-based Koopman modeling framework~\cite{desko,dlko_n,Li2024CCE,dlko} has emerged as a promising solution. Within this framework, the lifting functions can be automatically determined by training neural networks on batch system data~\cite{xiao2022deep}. Specifically, a neural network is established to map the original system state $x$ to a finite-dimensional lifted state space, where Koopman operators can be established via either solving the associated optimization directly~\cite{desko,dlko_n,Li2024CCE} or using the least-squares solution~\cite{dlko} as adapted from DMD and EDMD-based Koopman modeling approaches~\cite{edmd,dmd}. 

Compared to the conventional methods using manual or semi-automatic lifting function selection for Koopman modeling, the machine learning-based approaches enable automatic selection of the lifting functions from data. This eliminates the necessity of manually specifying the lifting functions or associated coefficients; consequently, this type of solution reduces the reliance of Koopman modeling on trial-and-error tests and users' experience, and can improve the applicability of Koopman modeling to practical systems of medium- to large-scales. Meanwhile, we note that the existing learning-based Koopman modeling approaches are purely data-driven~\cite{desko,dlko_n,Li2024CCE, dlko}.
This presents limitations in the following manner: 1) Effective training of neural networks requires a substantial amount of data to elucidate features effectively~\cite{bartlett2003vapnik,cho2015much}. With insufficient data samples, it is possible that a neural network model solely focuses on the features present in the training set, which may lead to overfitting~\cite{ying2019overview}. This issue is particularly critical for systems with a large number of state variables and when more complex neural networks with numerous layers and training parameters are employed. 2) Noise present in datasets can significantly disrupt feature extraction, which further leads to degraded modeling performance~\cite{probabilistic}. On the other hand, even partial information about $f$ in (\ref{paper1:model:nonlinear}) can contain valuable insights related to the real system dynamics. Integrating the available physical information can guide neural network training to prioritize extraction of features that are associated with system behaviors, thereby mitigating the influence of noise and outliers on modeling performance. This type of solution can generate more robust models even with small-scale and low-quality datasets.

Based on the above observations, we aim to propose a data-efficient machine learning-enabled Koopman modeling method
that effectively utilizes both the limited data and the available physical information.

\subsection{Proposed Koopman model structure}
\label{sec: model structure}

\begin{figure}[t]
    \centering
    \includegraphics[width=1\textwidth]{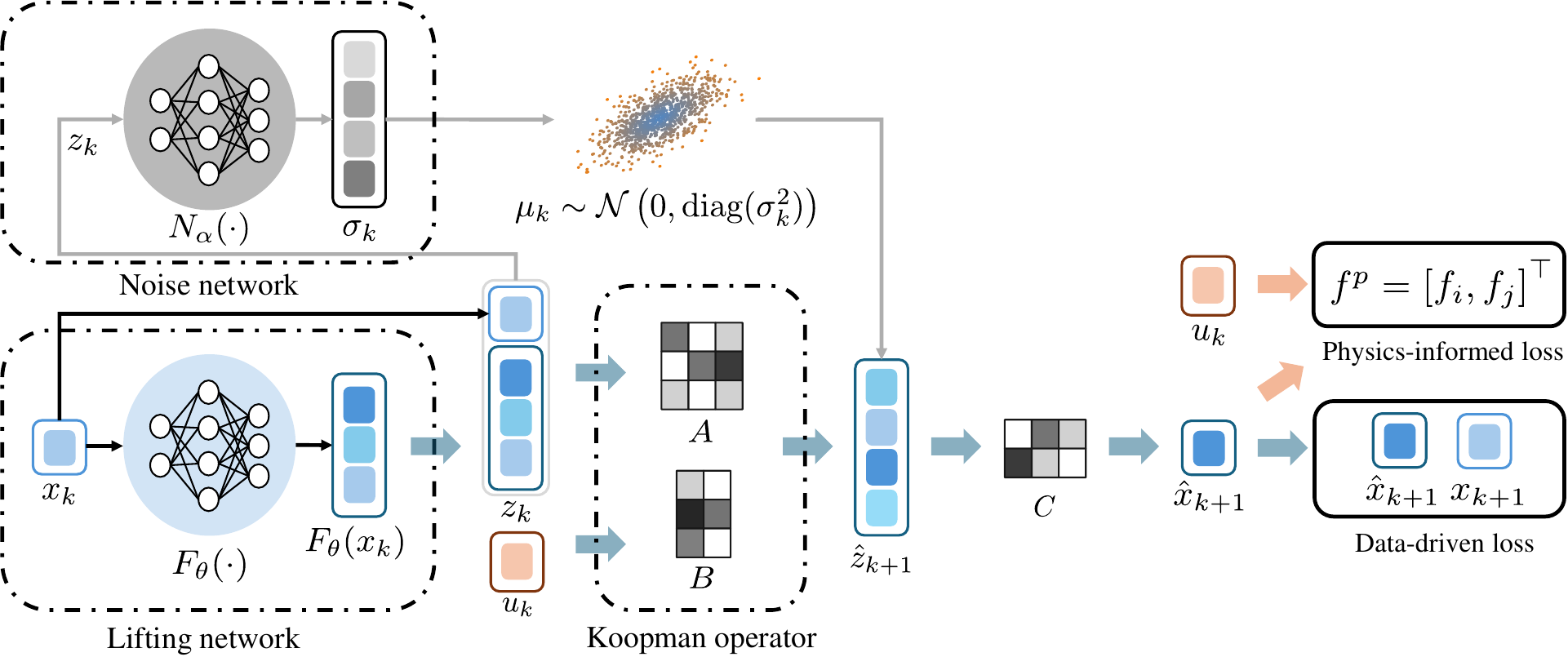}
    \caption{An overview of the proposed Koopman modeling approach. }
    \label{paper1:fig:model}
\end{figure}

In this section, we present the structure of the Koopman model. Implementing the proposed physics-informed Koopman modeling approach establishes a state lifting network $F_\theta(x)$, a noise characterization network $N_{\alpha}(z)$, and Koopman operators in the form of matrices $A$, $B$, and $C$, as shown in Figure~\ref{paper1:fig:model}. 
The two neural networks and the Koopman operators are integrated to form a stochastic Koopman model for multi-step ahead prediction of state $z$ in the lifted state space. Specifically, the stochastic Koopman model is in the following form:
\begin{subequations}
    \label{paper1:k:forward}
    \begin{align}
    \hat{z}_{k+j+1|k} &= A\hat{z}_{k+j|k} + Bu_{k+j}+ \mu_{k},\  j=0,\dots,H-1 \label{paper1:k:forward:1}\\
    \hat{x}_{k+j|k} &= C\hat{z}_{k+j|k} \label{paper1:k:forward:2}\\
    \hat y_{k+j|k} &= D\hat{z}_{k+j|k} \label{paper1:k:forward:3}
    \end{align}
\end{subequations}
where $H$ is the size of the prediction horizon; $\hat{z}_{k+j|k}$ is a prediction of $z_{k+j}$ obtained at time instant $k$; $\mu_k$ represents stochastic system disturbances with the prediction window starting from time instant $k$. We assume that $\mu_k$ follows a Gaussian distribution with varying standard deviation, which is similar to the setting in~\cite{probabilistic}, and we utilize the noise characterization network $N_\alpha(\cdot)$ to approximate the standard deviation for the time window starting from time instant $k$. The lifting functions are designed to be in the following form:

\begin{equation}
\label{paper1:z}
    z_k = g(x_k)= \left[
 \begin{array}{c}
 x_k\\
 F_\theta(x_k)
 \end{array}
 \right]
\end{equation}
such that the first $n_x$ elements of the lifted state are identical to the original state of system~(\ref{paper1:K:nonliear}), which facilitates the state estimation design. In~(\ref{paper1:z}), the state lifting network $F_\theta(x)$ accounts for the remaining $n_l$ elements of the lifting functions, where $n_l$ is the dimension of the output of the state lifting network.

At each sampling instant, $\mu_k$ is determined based on the output of the noise characterization network $N_\alpha(z_k)$. The Koopman model in~(\ref{paper1:k:forward}) generates multi-step ahead open-loop predictions of the lifted state. We use the reconstruction matrix $C$ to reconstruct the original system state based on the predictions of the lifted states using~(\ref{paper1:k:forward:2}). Trainable parameters associated with the Koopman model in~(\ref{paper1:k:forward}) include the parameters of the two neural networks $F_\theta(\cdot)$ and $N_\alpha(\cdot)$, and the Koopman operator matrices $A$ and $B$. In practical applications, matrices $C$ and $D$ can be determined without training. The determination of these two matrices will be explained in detail as we elaborate on the construction of the Koopman operator matrices.

The three primary components of the proposed Koopman model structure are detailed as follows:

\begin{itemize}
    \item \textbf{State lifting network $ F_{\theta} (\cdot)$}: The input to this multi-layer neural network is the original system state $x$. This network can be determined as a multi-layer neural network, with appropriately selected activation functions (e.g., ReLU, ELU, Tanh). This neural network encodes $x$ and generates a vector $F_\theta(x_{k})\in \mathbb{R}^{n_l}$. Then we concatenate the original state variables $x_{k}$ with the output of the neural network $F_\theta(x_{k})$ to obtain lifted state $z_k\in\mathbb{R}^{n_x} \to \mathbb{R}^{n_g}$.

    \item \textbf{Noise characterization network $N_\alpha(\cdot)$}: We assume that the noise in $\hat{\mathcal{G}}$ space follows the Gaussian distribution with zero mean and a time-varying standard deviation. This noise characterization network is used to provide an approximated standard deviation $\sigma_k \in \mathbb{R}^{n_g}$ of the noise distribution at each time instant $k$. The network takes lifted state $z_k$ as the input, and the output of this network is the logarithm of $\sigma_k$. Consequently, the standard deviation is calculated as follows:
    \begin{equation}
        \label{paper1:sigma}
        \sigma_k = e^{N_{\alpha}(z_k)}
    \end{equation}
    This ensures that the inferred standard deviation remains positive. 
    
    In each open-loop prediction window, the states $\hat{z}_{k+j|k}$, $j=1,\dots, H$, are generated through forward propagation using matrices $A$ and $B$ and the initial state $z_k$. All the state predictions generated within this multi-step ahead prediction window are generated based on the initial state $z_k$. Therefore, we use the first state of each window $z_k$ as the input of $N_\alpha(\cdot)$, noise $\mu_k$ remains unchanged within each multi-step ahead prediction window of size $H$.

    \item \textbf{Koopman operator}: Two matrices $A\in \mathbb{R}^{n_g\times n_g}$ and $B\in \mathbb{R}^{n_g\times n_u}$ need to be established to forecast the dynamic behavior of nonlinear system~(\ref{paper1:model:nonlinear}) in the higher-dimensional linear state-space. We use the reconstruction matrix $C\in \mathbb{R}^{n_x\times n_g}$ to reconstruct the original state $x$ using the prediction of the lifted state. It is not necessary to train matrices $C$ and $D$. Specifically, since $z_k$ is defined in a way that its first $n_x$ elements are identical to $x_k$, the reconstruction matrix $C$ can be determined as $C = [I_{n_x}, \bf{0}]$. In addition, due to the nature of most systems, measurements $y$ determined in~(\ref{paper1:model:Linear:3}) are linearly dependent on the original system state, that is, the output measurement function $h(x)$ in~(\ref{paper1:model:nonlinear:2}) is with a linear form of $h(x) = \bar C x$ where $\bar C \in \mathbb R^{n_y \times n_x}$ is a matrix pre-determined according to the physical meanings of the sensor measurements. In this case, matrix $D$ in~(\ref{paper1:k:forward:3}) is with the form of $D = \bar C C$. 
    
\end{itemize}

\begin{rmk}
In Figure~\ref{paper1:fig:model}, $\mathcal{L}_{data}$ contains data-driven penalties on the model prediction errors, and $\mathcal{L}_{PI}$ contains physics-based penalties on the prediction errors. Training a stable noise characterization network subject to random initialization can be challenging, mainly due to the equivalence between physical information and information contained in noise-free data. To enhance training stability, we pre-train a noise characterization network using the pipeline shown in Figure~\ref{paper1:fig:model} without physical information. Throughout the training of the physics-informed stochastic Koopman model, the parameters $\alpha$ of $N_\alpha(\cdot)$ remain unchanged.
\end{rmk}

\begin{rmk}
In this work, we build a stochastic Koopman model for two primary reasons: 1) to handle unknown disturbances to the underlying nonlinear system; 2) to account for the mismatch between the actual system and the Koopman model. Additionally, introducing stochasticity into the Koopman model structure aligns with the need to design MHE schemes.
\end{rmk}

\subsection{Physics-informed Koopman model training}
\label{sec:pi optimal}

The implementation of the proposed method utilizes the known physical information $f^p$ and a collected dataset contains $N+H$ samples, where $N$ represents the number of trajectories of length $H$ that can be formed using the samples from this dataset.
As depicted in Figure~\ref{paper1:fig:model}, the proposed method evaluates and optimizes the predictive performance by respecting both ground-truth data and the available physical information -- the state predictions given by the Koopman model are deemed sufficiently accurate only when they align with the limited data while remaining consistent with the available first-principles equations as part of the vector function $f$ in~(\ref{paper1:model:nonlinear:1}). Therefore, the formulated optimization problem should include both the data-driven loss functions and the physics-informed loss functions.

The Koopman model needs to be trained in a way such that it has multi-step ahead predictive capability within a prediction horizon of size $H$ as described in~(\ref{paper1:k:forward}). Specifically, the Koopman model training phase focuses on four objectives: 1) to ensure the state prediction accurately approximates the ground truth; 2) to ensure the prediction of the lifted state accurately approximates the output of actual lifted state in the lifted space $\hat{\mathcal{G}}$; 3) to align the physics-related part of the predicted state sequence with the available physical information contained in $f^p$; 4) to ensure that the sequence of the predicted lifted state aligns with the lifted state sequence inferred from physical information.

Consequently, the optimization problem associated with Koopman modeling is formulated as follows:
\begin{subequations}
    \label{paper1:Loss:total}
    \begin{align}
    \min_{\underset{\theta,\alpha, A, B}{}}\ & \epsilon_1\mathcal{L}_x + \epsilon_2\mathcal{L}_z + \epsilon_3\mathcal{L}_{px} + \epsilon_4\mathcal{L}_{pz} \label{paper1:Loss:total:1}\\
    \text{s.t.}\  \hat{z}_{k+j+1|k} &= A\hat{z}_{k+j|k} + Bu_{k+j} + \mu_k,\  j=0,\dots,H-1 \label{paper1:Loss:total:constraints}\\
    \hat{x}_{k+j|k} &= C\hat{z}_{k+j|k}\\
    \hat{z}_{k|k} &= g(x_k)\\
    \mu_k &\sim \mathcal{N}\left(0,\text{diag}(\sigma_k^{2})\right)
    \end{align}
\end{subequations}
where $\theta$ denotes the parameters of the state lifting network; $\alpha$ denotes the parameters of the noise characterization network. $\mathcal{L}_{x}$, $\mathcal{L}_{z}$, $\mathcal{L}_{px}$, and $\mathcal{L}_{pz}$ represent the loss functions corresponding to the four tasks mentioned above, and will be represented in detail. $\epsilon_i$, $i=1,2,3,4$, denote the adaptive weights that are updated after every epoch. The first two terms consider only data while the other two terms are physics-informed loss terms. We aim for the model to have multi-step ahead prediction capability and minimize the cumulative error within each prediction horizon of $H$ steps. The four terms on the right-hand side of~(\ref{paper1:Loss:total:1}) take into account multi-step ahead prediction errors for the original state $x$ and the lifted state $z$, respectively.

In (\ref{paper1:Loss:total:1}), the first two terms $\mathcal{L}_x$ and $\mathcal{L}_z$ are the data-driven penalties on the sum of the state prediction loss in the original space and the sum of the linear propagation loss in the lifted space, respectively:
\begin{subequations}
    \label{paper1:Loss:d}
    \begin{align}
    \mathcal{L}_x &= \frac{1}{NH}\sum_{k=0}^N\sum_{j=0}^H||x_{k+j} - C\hat{z}_{k+j|k}||^2 \label{paper1:Loss:d:2}\\
    \mathcal{L}_z &= \frac{1}{NH}\sum_{k=0}^N\sum_{j=0}^H||z_{k+j} - \hat{z}_{k+j|k}||^2 \label{paper1:Loss:d:3}
    \end{align}
\end{subequations}
In (\ref{paper1:Loss:d}), the predictions of the original state and the lifted state are considered. In~(\ref{paper1:Loss:d:2}), $C\hat{z}$ represents the reconstructed system state according to~(\ref{paper1:k:forward:2}); $x$ is the ground-truth data of the original system state, contained in the dataset. 
In~(\ref{paper1:Loss:d:3}), $\hat{z}_{k+1:k+H|k}$ is generated following~(\ref{paper1:k:forward:1}) subject to the initial state $\hat{z}_{k|k}$, Koopman operators $A$ and $B$, and the stochastic disturbance $\mu_k$ as included in~(\ref{paper1:k:forward:1}); $z_{k+j}$ is generated by inputting $x_{k+j}$ to the lifting function, that is, $z_{k+j}=g(x_{k+j})$.

In addition to data-driven penalties, we also use partially available information about $f$, denoted by $f^p$, 
to formulate physics-based penalties on the sum of original state prediction loss and the sum of linear propagation loss in the lifted space, denoted by $\mathcal{L}_{px}$ and $\mathcal{L}_{pz}$ in (\ref{paper1:Loss:total:1}), respectively. In the original state space, it is reasonable to expect that the reconstructed state $\hat{x}$ should conform to the available physical information $f^p$, as expressed in~(\ref{paper1:Loss:pi:1}). 
\begin{subequations}
    \label{paper1:Loss:pi:1}
    \begin{align}
    \mathcal{L}_{px} =& \frac{1}{NH}\sum_{k=0}^N\sum_{j=0}^{H-1}||\bar{x}_{k+j+1|k}^p-\hat{x}_{k+j+1|k}^p||^2 \label{paper1:loss:pi:11}\\
    \bar{x}_{k+j+1|k}^p&= f^p(\hat{x}_{k+j|k}^p, u_{k+j}),\ j=0,\dots,H-1 \label{paper1:loss:pi:12}
    \end{align}
\end{subequations}
where $\hat{x}_{k+j|k}^p$ refers to the reconstructed states that are associated with the available physical information; $\bar{x}_{k+j+1|k}^p$ is a one-step ahead open-loop prediction obtained using the available physical knowledge. 
$\hat{x}^p_{k+j|k}$ and $\hat{x}^p_{k+j+1|k}$, $ j=0,\dots, H-1$, which are corresponding to two consecutive sampling instants, should adhere to the system dynamics $f^p$.
Since we only know partial information about $f$, we select $\hat{x}^p$ which corresponds to $f^p$ in the physics-informed penalty on the original state prediction error. As demonstrated by~(\ref{paper1:loss:pi:12}), we substitute the current predicted state $\hat{x}^p_{k+j|k}$ and system input $u_{k+j}$ into $f^p$ to obtain the next estimated state $\bar{x}_{k+j+1|k}^p$.

Moreover, we leverage the available physical information to penalize the difference between the predicted lifted state $\hat{z}_{k+j|k}$ and the lifted state that is inferred from physical equations $\bar{z}_{k+j|k}$ using $\mathcal{L}_{pz}$ as follows:
\begin{subequations}
    \label{paper1:Loss:pi:2}
    \begin{align}
    &\mathcal{L}_{pz} = \frac{1}{NH}\sum_{k=0}^N\sum_{j=0}^{H-1}||\bar{z}_{k+j+1|k}-\hat{z}_{k+j+1|k}||^2 \label{paper1:Loss:pi:21}\\
    &\bar{z}_{k+j+1|k} = g(\bar{x}_{k+j+1|k}),\ j=0,\dots,H-1\label{paper1:Loss:pi:22}
    \end{align}
\end{subequations} 
where $\bar{x}_{k+j+1|k}$ is formed by incorporating $\bar{x}_{k+j+1|k}^p$, while the remaining components are taken from $\hat{x}_{k+j+1|k}$.
Because the state lifting network $F_\theta(\cdot)$ has a fixed network structure, the input vector must have the same dimension as the original system state $x$.


\subsection{Weights adaptation using maximum likelihood estimation}
\label{sec: mle weight}
The predictive capability of the established Koopman model can be significantly influenced by the weights $\epsilon_i$, $i=1,2,3,4$, for the four terms on the right-hand side of~(\ref{paper1:Loss:total}). We introduce an adaptive parameter optimization method, which is based on maximum likelihood estimation, to optimize the weights in (\ref{paper1:Loss:total:1}).

The error of a machine learning model varies across different tasks~\cite{modeluncertainties, mtl}. While~(\ref{paper1:Loss:total}) results in a single Koopman model, the four terms on the right-hand side of~(\ref{paper1:Loss:total:1}) can be considered four individual tasks. By assigning higher weights to tasks in which we have greater confidence while assigning smaller weights to tasks in which our confidence is lower, we can potentially enhance the modeling performance~\cite{mtl, mtlr}. 

First, let us consider the first task corresponding to $\mathcal L_x$ in (\ref{paper1:Loss:total:1}), which focuses on finding Koopman operator matrices $A$  and $B$ and lifting function $g(\cdot)$ that minimize the discrepancy between the predicted state and its corresponding ground-truth in the original state-space. 
Let $\eta_1$ represent the the task-dependent model uncertainty for this specific task and the following equation holds:
\begin{equation}
\label{paper1:mle:fw}
    x_{k+1} = C(Ag(x_{k})+Bu_k) + \eta_1.
\end{equation}

We assume this uncertainty follows a Gaussian distribution with zero mean and standard deviation $\nu_1$, that is, $\eta_1 \sim \mathcal{N}(0, \nu_1^2)$. In this case, the probability distribution of the ground-truth $x_{k+1}$ follows a Gaussian distribution with the mean value $ C(Ag(x_{k})+Bu_k) $ and standard deviation $\nu_1$ in the following form:
\begin{equation}\label{Nofl_x}
    x_{k+1} \sim P\big(x_{k+1}|C(Ag(x_{k})+Bu_k)\big) = \mathcal{N}\big(C(Ag(x_{k})+Bu_k),\nu_1^2\big)= \frac{1}{\nu_1\sqrt{2\pi}}e^{-\frac{\left(x_{k+1}-C(Ag(x_{k})+Bu_k)\right)^2}{2\nu_1^2}}
\end{equation}

We make similar assumptions on the distributions of the model uncertainties for the remaining three tasks, and let $\nu_i,i=2,3,4$, represent the standard deviations of the task-dependent modeling model uncertainties for the remaining three tasks in~(\ref{paper1:Loss:total:1}). The following equations hold, which focus on finding Koopman operator matrices $A$ and $B$ and lifting function $g(\cdot)$ that minimize the discrepancy between the predicted state and its corresponding ground-truth in the original state-space:
\begin{subequations}
\label{paper1:mle:distribution}
\begin{align}
     z_{k+1} &\sim P\big(z_{k+1}|Ag(x_k)+Bu_k\big) = \frac{1}{\nu_2\sqrt{2\pi}}e^{-\frac{\left(z_{k+1}-(Ag(x_k)+Bu_k)\right)^2}{2\nu_2^2}}\label{paper1:mle:distribution:2}\\
     \bar x_{k+1}^p &\sim P\left(\bar{x}_{k+1}^p|f^p\big(C(Ag(x_{k-1})+Bu_{k-1})\big)\right) = \frac{1}{\nu_3\sqrt{2\pi}}e^{-\frac{\left(\bar{x}_{k+1}^p-f^p\left(C(Ag(x_{k-1})+Bu_{k-1})\right)\right)^2}{2\nu_3^2}}\label{paper1:mle:distribution:3}\\
     \bar z_{k+1} &\sim P\Big(\bar z_{k+1}|g\big(\Tilde f\left(C\left(Ag(x_{k-1})+Bu_{k-1}\right)\right)\big)\Big)= \frac{1}{\nu_4\sqrt{2\pi}}e^{-\frac{\Big(\bar z_{k+1}-g\big(\Tilde f\left(C\left(Ag(x_{k-1})+Bu_{k-1}\right)\right)\big)\Big)^2}{2\nu_4^2}}\label{paper1:mle:distribution:4}
\end{align}
\end{subequations}
where $\Tilde{f}\left(C\left(Ag(x_{k-1})+Bu_{k-1}\right)\right)$ is formed by integrating the one-step ahead prediction obtained using $f^p$ with the two-step ahead prediction obtained using the Koopman model in~(\ref{paper1:k:forward}).

Let $y_1$, $y_2$, $y_3$, and $y_4$ represent outputs $x_{k+1}$, $z_{k+1}$, ${\bar{x}}_{k+1}^p$, and ${\bar{z}}_{k+1}$ respectively, and let $W$ represent the set of trainable parameters $W = \{g(\cdot),A,B\}$. Based on the principle of maximum likelihood estimation, considering the final loss function in (\ref{paper1:Loss:total:1}) comprising four tasks, we aim to maximize the joint probability of~(\ref{Nofl_x}) and (\ref{paper1:mle:distribution}) given as follows:
\begin{subequations}
\label{paper1:mle:max}
\begin{align}
    P\left(y_1,y_2,y_3,y_4|W\right)&= 
    P\big(y_1|C(Ag(x_{k})+Bu_k))P(y_2|Ag(x_k)+Bu_k\big)\times\label{paper1:mle:max:1}\\
    &P\left(y_3|f^p\big(C(Ag(x_{k-1})+Bu_{k-1})\big)\right)P\Big(y_4|g\big(\Tilde{f}\left(C\left(Ag(x_{k-1})+Bu_k\right)\right)\big)\Big)\notag\\
    &=\frac{1}{4\pi^2\prod_{i=1}^{4}\nu_i} e^{\gamma(k)}\label{paper1:mle:max:2}
\end{align} 
\end{subequations}
where $\frac{(y_1-C(Ag(x_{k})+Bu_k))^2}{2\nu_1^2} + \frac{(y_2-(Ag(x_k)+Bu_k))^2}{2\nu_2^2}+\frac{(y_3-f^p\left(C(Ag(x_{k}^p)+Bu_k)\right)^2}{2\nu_3^2} + \frac{(y_4-g(\Tilde{f}\left(C\left(Ag(x_{k-1})+Bu_{k-1}\right)\right)))^2}{2\nu_4^2}$ is the exponent $\gamma(k)$. 


Furthermore, due to the monotonic increasing property of a logarithmic function, maximizing~(\ref{paper1:mle:max}) is equivalent to maximizing its logarithm. Therefore, we take the logarithm of~(\ref{paper1:mle:max}) and discard the constant term $\frac{1}{2}\text{log}(2\pi)$ which does not include any trainable parameters. Note that in $\gamma(k)$, $y_1-C(Ag(x_{k})+Bu_k)$, $y_2-(Ag(x_k)+Bu_k)$, $y_3-f^p\big(C(Ag(x_{k-1})+Bu_{k-1})\big)$, and $y_4-g\big(\Tilde{f}\left(C\left(Ag(x_{k-1})+Bu_k\right)\right)\big)$ squared are equivalent to $\mathcal{L}_x$, $\mathcal{L}_z$, $\mathcal{L}_{px}$, and $\mathcal{L}_{pz}$ in (\ref{paper1:Loss:total:1}), respectively. We convert~(\ref{paper1:mle:max:1}) into a minimization problem by assigning negative signs and incorporating the mean squared error terms into~(\ref{paper1:mle:max:2}). This results in a function that takes the form of a weighted sum of multiple mean squared error losses along with a regularization term:
\begin{subequations}
\label{paper1:mle:min}
\begin{align}
    \mathcal{L} &= \frac{1}{2\nu_1^2}\mathcal{L}_x + \frac{1}{2\nu_2^2}\mathcal{L}_z + \frac{1}{2\nu_3^2}\mathcal{L}_{px} + \frac{1}{2\nu_4^2}\mathcal{L}_{pz} + \log\ \prod_{i=1}^{4} \nu_i\\
    &=\frac{1}{2\nu_1^2}\mathcal{L}_x + \frac{1}{2\nu_2^2}\mathcal{L}_z + \frac{1}{2\nu_3^2}\mathcal{L}_{px} + \frac{1}{2\nu_4^2}\mathcal{L}_{pz} + \sum_{i=1}^4 \log\nu_i
\end{align}
\end{subequations}

Moreover, we make two adjustments to the regularization term. First, we replace $\log \nu_i$ to $\log(\nu_i + 1)$, which ensures the regularization term is positive~\cite{mtlr}. Second, we introduce a scaling factor $\beta$ to avoid large $\nu_i$ values. This adjustment ensures that the reduction in overall loss is driven by the intended loss terms, rather than being dominated by large $\nu_i$ values.
Consequently, by including the adjusted regularization, the optimization problem for physics-informed Koopman modeling is formulated as follows:
\begin{equation}
\label{paper1:loss:final}
    \min_{W}\ \frac{1}{2\nu_1^2}\mathcal{L}_x + \frac{1}{2\nu_2^2}\mathcal{L}_z + \frac{1}{2\nu_3^2}\mathcal{L}_{px} + \frac{1}{2\nu_4^2}\mathcal{L}_{pz} + \beta\sum_{i=1}^4 \log(1+\nu_i).
\end{equation}

\section{Learning-based self-tuning MHE}
 
\label{sec:MHE}
\begin{figure}[t]
    \centering
    \includegraphics[width=1\textwidth]{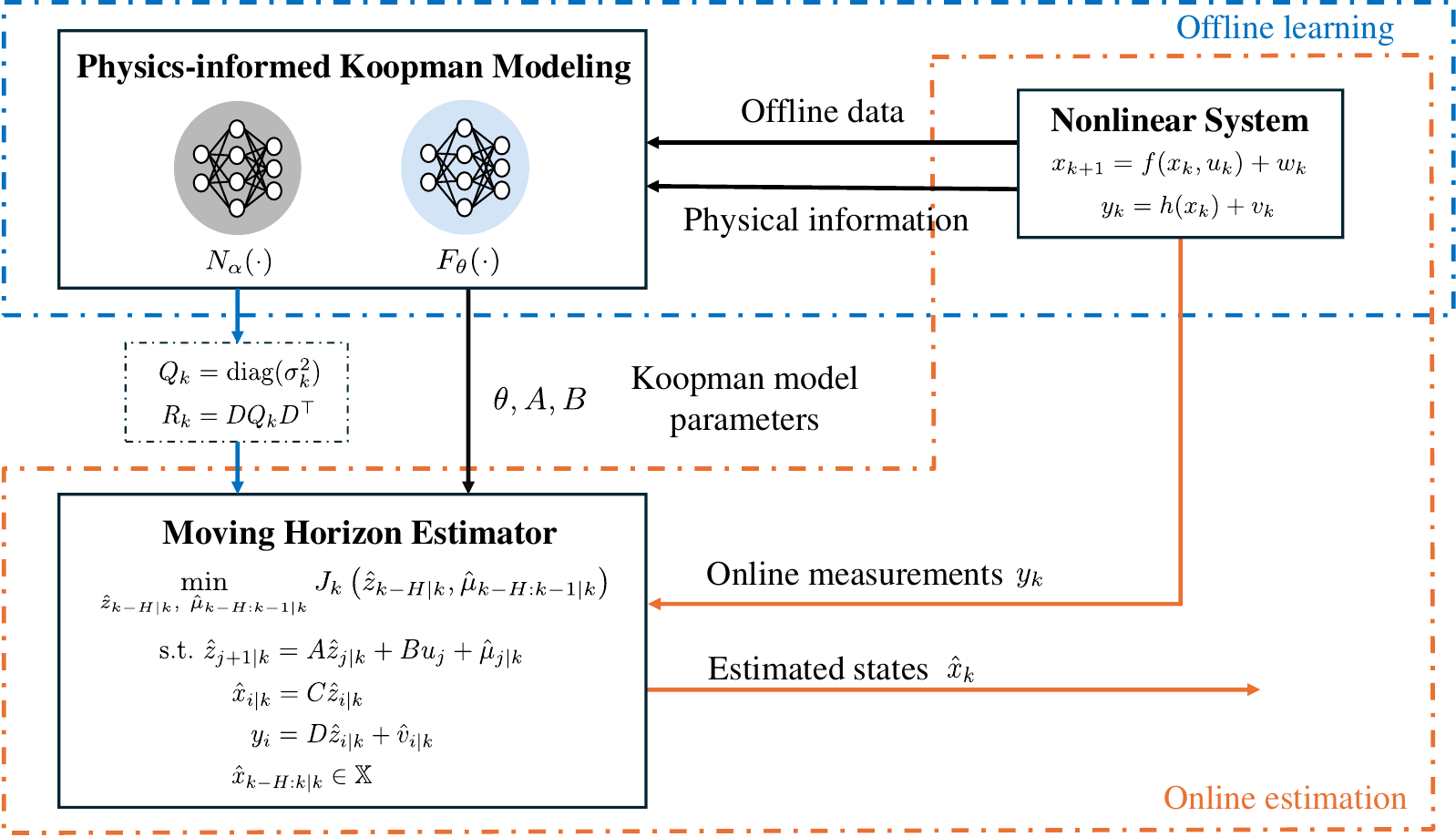}
    \caption{An overview of the proposed learning-based self-tuning Koopman MHE.}
    \label{fig:mhe}
\end{figure}
In this section, we develop a learning-based moving horizon estimation (MHE) method based on the established Koopman model to efficiently estimate the full state of the nonlinear system in~(\ref{paper1:model:nonlinear}) in real-time. In this method, we propose a self-tuning weighting matrix update strategy that updates the weighting matrices of the MHE-based estimator for improved estimation performance. The proposed learning-based self-tuning MHE and its connection to the offline physics-informed Koopman modeling is illustrated in Figure~\ref{fig:mhe}. In this figure, the blue solid lines indicate the parameters that are updated at each sampling instant, while the black lines represent the parameters that remain constant throughout the state estimation process.

At each sampling instant $k\geq H$, the self-tuning MHE estimator aims to find a sequence of optimal estimates $\hat{x}_{k-H|k},\dots,\hat{x}_{k|k}$ based on the output measurements within a time window of size $H$, that is, vector $y_{k-H:k}$. The optimization problem associated with the self-tuning MHE is as follows:
\begin{subequations}\label{paper1:mhe}
\begin{align}
        \min_{\underset{\hat{z}_{k-H|k},\ \hat{\mu}_{k-H:k-1|k}}{}} & J_k(\hat{z}_{k-H|k},\hat{\mu}_{k-H:k-1|k})\\
        \text{s.t.}\ \hat{z}_{j+1|k}&=A\hat{z}_{j|k}+Bu_j+\hat{\mu}_{j|k},\ j=k-H,\dots,k-1\label{paper1:mhe:linear}\\
        \hat{x}_{i|k} &= C\hat{z}_{i|k}\\
        y_{i} &= D\hat{z}_{i|k} + \hat{v}_{i|k},\ i=k-H,\dots,k\\
        \hat{x}_{k-H:k|k}&\in \mathbb{X}
\end{align}
\end{subequations}
where the objective function $J_k$ is adapted from~\cite{minmaxnonlinear} as follows:
\begin{equation}
    \label{paper1:mhe:obj}
    J_k(\hat{z}_{k-H|k},\hat{\mu}_{k-H:k-1|k}) = ||\hat{z}_{k-H|k}-\bar{z}_{k-H}||^2 + \sum_{j=k-H}^{k-1}l(\hat{\mu}_{j|k},\hat{v}_{j|k}) + \max_{\underset{j \in k-N:k-1}{}}l(\hat{\mu}_{j|k},\hat{v}_{j|k}) 
\end{equation}
with the stage cost being
\begin{equation}\label{paper1:mhe:l}
l(\hat{\mu}_{j|k},\hat{v}_{j|k}) = ||\hat{\mu}_{j|k}||_{Q_k^{-1}}^{2}+||\hat{v}_{j|k}||_{R_k^{-1}}^{2}
\end{equation}

In~(\ref{paper1:mhe})-(\ref{paper1:mhe:l}), $\hat{z}_{j|k}$, $j = k-H,\dots, k$, is the estimate of the lifted state; $\hat{\mu}_{k-H:k-1|k}$ is the sequence of the estimated the disturbances in the lifted state-space; $\hat{v}_{j|k}$ is an estimate of the mismatch of the output measurement equation in the lifted space for time instant $j$ obtained at time $k$.
The first term on the right-hand side of~(\ref{paper1:mhe:obj}) is the arrival cost that summarizes the historical information prior to the current estimation window~\cite{robustmhe}, where $\bar{z}_{k-H}$ is the $\emph{a priori}$ estimate of $z_{k-H}$ computed following $\bar{z}_{k-H} = A\hat{z}_{k-H-1|k-1} + Bu_{k-H-1}$. The third term on the right-hand side of (\ref{paper1:mhe:obj}) plays an important role in improving the robustness of the estimates~\cite{minmaxmhe,minmaxnonlinear}. $Q_k$ and $R_k$ in~(\ref{paper1:mhe:l}) are two weighting matrices; they are updated at each sampling instant $k$ based on the output of the pre-trained noise characterization network $N_{\alpha}(\cdot)$. As described in Figure~\ref{fig:mhe}, at each sampling instant $k$, $N_\alpha(\bar{z}_{k-H})$ outputs the logarithms of $\sigma_k$ which are the estimate of the time-varying standard deviation of the disturbance $\mu_k$ within the lifted space. $\sigma_k$ are used to update weight matrices $Q_k$ as follows:
\begin{equation}
    Q_k = \text{diag}(\sigma_k^{2})\label{paper1:mhe:Q}
\end{equation}

$R_k$, which accounts for the variance of the measurement noise, is determined based on $Q_k$ as:
\begin{equation}
    R_k = D Q_k D^{\top}\label{paper1:mhe:R}
\end{equation}

\begin{rmk}
To update $Q$ and $R$, the input to $N_{\alpha}(\cdot)$ is the initial prediction $\bar{z}_{k-H}$ generated at the previous stage $k-1$ since the only available full state information is the estimation obtained at the previous step. Additionally, during state estimation implementation, we scale the data to have a mean of 0 and a standard deviation of 1. Accordingly, $Q_k$ is also scaled.
\end{rmk}

\section{Application to a chemical process}
\label{sec:case}
In this section, a benchmark chemical process with bounded random system disturbances is utilized to illustrate the efficacy and superiority of the proposed physics-informed Koopman modeling and self-tuning MHE. 
\subsection{Process description}
\begin{figure}[t]
    \centering
    \includegraphics[width= 0.7\textwidth]{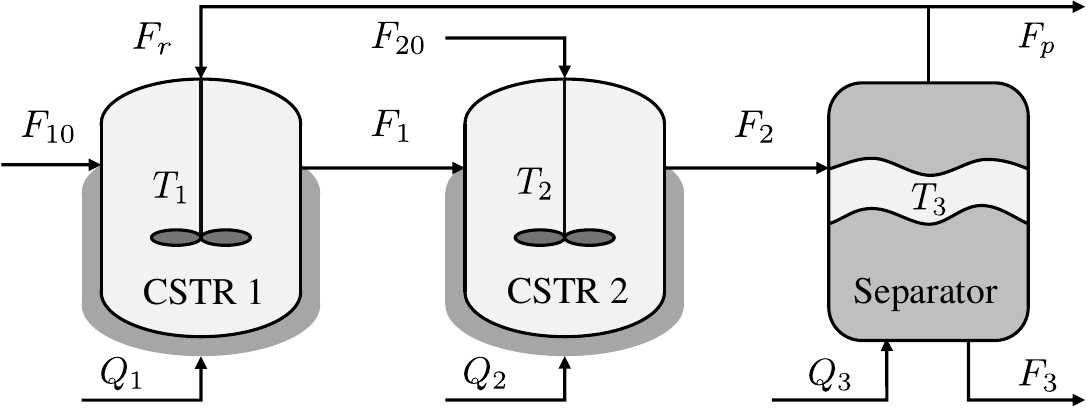}
    \caption{A schematic of the reactor-separator chemical process.} 
    \label{fig:cstr}
\end{figure}
This chemical process involves two continuously stirred tank reactors (CSTRs) and one flash tank separator. A schematic of this process is shown in Figure~\ref{fig:cstr}. In this process, reactant $\text{A}$ is converted into desired product $\text{B}$, while a portion of $\text{B}$ is further transformed into side product $\text{C}$. The two chemical reactions characterized as $\text{A}\to \text{B}$ and $\text{B} \to \text{C}$ take place simultaneously within the two reactors, referred to as CSTR 1 and CSTR 2, as shown in Figure~\ref{fig:cstr}. A fresh feed flow containing reactant $\text{A}$ is introduced into CSTR 1 at flow rate $F_{10}$ and temperature $T_{10}$. The effluent of CSTR 1 enters CSTR 2 at flow rate $F_1$ and temperature $T_1$, while another feed stream carrying pure $\text{A}$ enters CSTR 2 at flow rate $F_{20}$ and temperature $T_{20}$. The effluent of CSTR 2 is directed to the separator at flow rate $F_2$ and temperature $T_2$. The separator has a recycle stream back to the first vessel at flow rate $F_r$ and temperature $T_3$. Each of the three vessels has a heating jacket, which can add heat to/remove heat from the corresponding vessel at heating input rate, $Q_i,~i=1,2,3$, respectively. The state vector $x$ of this process encompasses nine state variables, specifically, $x = [x_{\text{A}1},x_{\text{B}1},T_1,x_{\text{A}2},x_{\text{B}2},T_2,x_{\text{A}3},x_{\text{B}3},T_3]^\top$, where $x_{\text{A}i}$ and $x_{\text{B}i}$, denote the mass fractions of $\text{A}$ and $\text{B}$ in the $i$th vessel; $T_i$ represents the temperature in the $i$th vessel, $i=1,2,3$. Nine ordinary differential equations (ODEs) are established to describe the process dynamic behaviors based on material and energy balances~\cite{cstr}:
\begin{subequations}\label{paper1: cstr:equation}
\begin{align}
& \displaystyle{\frac{dx_{\text{A}1}}{dt}} = \frac{F_{10}}{V_1}(x_{\text{A}10} - x_{\text{A}1}) + \frac{F_r}{V_1}(x_{\text{A}r} - x_{\text{A}1}) - k_1 e^{\frac{-E_1}{RT_1 }}x_{\text{A}1} \label{paper1: cstr:equation:1} \\[0.3em]
& \displaystyle{\frac{dx_{\text{B}1}}{dt}} = \frac{F_{10}}{V_1}(x_{\text{B}10} - x_{\text{B}1}) +  \frac{F_r}{V_1}(x_{\text{B}r} - x_{\text{B}1})+  k_1 e^{\frac{-E_1}{RT_1 }}x_{\text{A}1} - k_2 e^{\frac{-E_2}{RT_1 }}x_{\text{B}1} \label{paper1: cstr:equation:2} \\[0.3em]
& \displaystyle{~~\frac{dT_{1}}{dt}} = \frac{F_{10}}{V_1}(T_{10} - T_{1}) + \frac{F_r}{V_1}(T_{3} - T_{1}) - \frac{\Delta H_1}{ c_p}k_{1}e^{\frac{-E_1}{RT_1}}x_{\text{A}1} - \frac{\Delta H_2}{ c_p}k_{2}e^{\frac{-E_2}{RT_1}}x_{\text{B}1} +  \frac{Q_1}{\rho c_pV_1} \label{paper1: cstr:equation:3}\\[0.3em]
& \displaystyle{\frac{dx_{\text{A}2}}{dt}} = \frac{F_{1}}{V_2}(x_{\text{A}1} - x_{\text{A}2}) + \frac{F_{20}}{V_2}(x_{\text{A}20} - x_{\text{A}2}) - k_1 e^{\frac{-E_1}{RT_2 }}x_{\text{A}2} \label{paper1: cstr:equation:4} \\[0.3em]
& \displaystyle{\frac{dx_{\text{B}2}}{dt}} = \frac{F_{1}}{V_2}(x_{\text{B}1} - x_{\text{B}2}) + \frac{F_{20}}{V_2}(x_{\text{B}20} - x_{\text{B}2})+  k_1 e^{\frac{-E_1}{RT_2 }}x_{\text{A}2} - k_2 e^{\frac{-E_2}{RT_2}}x_{\text{B}2} \label{paper1: cstr:equation:5} \\[0.3em]
& \displaystyle{~~\frac{dT_{2}}{dt}} = \frac{F_{1}}{V_2}(T_{1} - T_{2}) + \frac{F_{20}}{V_2}(T_{20} - T_{2}) - \frac{\Delta H_1}{ c_p}k_{1}e^{\frac{-E_1}{RT_2}}x_{\text{A}2} - \frac{\Delta H_2}{ c_p}k_{2}e^{\frac{-E_2}{RT_2}}x_{\text{B}2} +  \frac{Q_2}{\rho c_pV_2} \label{paper1: cstr:equation:6}\\[0.3em]
& \displaystyle{\frac{dx_{\text{A}3}}{dt}} = \frac{F_{2}}{V_3}(x_{\text{A}2} - x_{\text{A}3}) - \frac{(F_r+F_p)}{V_3}(x_{\text{A}r} - x_{\text{A}3}) \label{paper1: cstr:equation:7} \\[0.3em]
& \displaystyle{\frac{dx_{\text{B}3}}{dt}} = \frac{F_{2}}{V_3}(x_{\text{B}2} - x_{\text{B}3}) - \frac{(F_r+F_p)}{V_3}(x_{\text{B}r} - x_{\text{B}3}) \label{paper1: cstr:equation:8} \\[0.3em]
& \displaystyle{~~\frac{dT_{3}}{dt}} = \frac{F_{2}}{V_3}(T_{2} - T_{3}) + \frac{Q_3}{\rho c_pV_3}+\frac{(F_{r}+F_{p})}{\rho c_{p}V_{3}}(x_{\text{A}r}\Delta H_{\text{vap1}}+x_{\text{B}r}\Delta H_{\text{vap2}}+x_{\text{C}r}\Delta H_{\text{vap3}} ) \label{paper1: cstr:equation:9}
\end{align}
\end{subequations}
where $x_{\text{A}10}$, $x_{\text{B}10}$ are the mass fractions of $\text{A}$ and $\text{B}$ in the feed flow; $x_{\text{A}r}$, $x_{\text{B}r}$, $x_{\text{C}r}$ are the mass fractions of $\text{A}$, $\text{B}$, $\text{C}$ in the recycle flow; $F_1$ and $F_2$ represent the effluent flow rates from CSTR 1 and CSTR 2, respectively; $F_r$, $F_p$ are the flow rates of the recycle flow and purge flow, respectively; $V_1$, $V_2$, $V_3$ are the volumes of the three vessels; $E_1$, $E_2$ are the activation energy for the two reactions; $k_1$, $k_2$ are the pre-exponential values for the two reactions; $\Delta H_1$, $\Delta H_2$ are the reaction heats for the two reactions; $c_p$ represents the heat capacity; $r$ is the gas constant; $\rho$ is the solution density.

Furthermore, we use $x_{\text{C}3}$ to represent the mass fraction of $\text{C}$ in the separator, and the algebraic equations describing the relationship between the composition of the overhead stream and the liquid composition in the separator are given as follows:
\begin{subequations}\label{paper1: cstr: algequ}
\begin{align}
x_{\text{A}r} &= \frac{\alpha _\text{A} x_{\text{A}3}}{\alpha _\text{A} x_{\text{A}3} + \alpha _\text{B} x_{\text{B}3} + \alpha _\text{C} x_{\text{C}3} } \\[0.3em]
x_{\text{B}r} &= \frac{\alpha _\text{B} x_{\text{B}3}}{\alpha _\text{A} x_{\text{A}3} + \alpha _\text{B} x_{\text{B}3} + \alpha _\text{C} x_{\text{C}3} } \\[0.3em]
x_{\text{C}r} &= \frac{\alpha _\text{C} x_{\text{C}3}}{\alpha _\text{A} x_{\text{A}3} + \alpha _\text{B} x_{\text{B}3} + \alpha _\text{C} x_{\text{C}3} } 
\end{align}
\end{subequations}

More detailed process descriptions and the process parameters used in this set of simulations can be found in~\cite{cstr}. The dynamic model in~(\ref{paper1: cstr:equation}) is used as a simulator for data generation. Additionally, in this work, we consider practical scenarios when only partial equations of (\ref{paper1: cstr:equation}) are available, which will be detailed in the subsequent subsection. The available physical information and the simulated data are jointly used for building a Koopman model that describes the dynamics of the process in a linear manner. 

\subsection{Simulation settings}

\begin{figure}[t]
    \centering
    \includegraphics[width=0.8\textwidth]{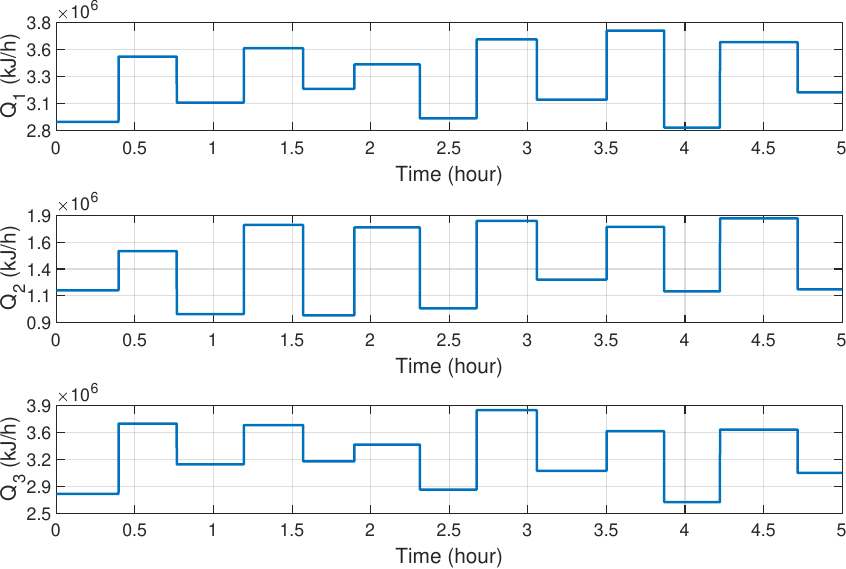}
    \caption{One set of open-loop control input trajectories used for data generation.}
    \label{fig:input}
\end{figure}

To illustrate the proposed modeling approach, we consider that among (\ref{paper1: cstr:equation}), only the ODEs that describe the dynamic behaviors of the temperatures in the three vessels are known. Specifically, only~(\ref{paper1: cstr:equation:3}), (\ref{paper1: cstr:equation:6}), and (\ref{paper1: cstr:equation:9}) are available, while the remaining ODEs are unavailable. Consequently, $f^p = [f_3, f_6, f_9]^{\top}$, where $f_3$, $f_6$, $f_9$ are the right-hand sides of (\ref{paper1: cstr:equation:3}), (\ref{paper1: cstr:equation:6}), and (\ref{paper1: cstr:equation:9}), respectively. The available physical information $f^p$ is leveraged for learning-based physics-informed Koopman modeling.

We consider a steady-state point $x_s = [0.1763, 0.6731,$ \allowbreak $ 480.3165~\text{K},$ \allowbreak $0.1965, 0.6536, 472.7863~\text{K},$ \allowbreak $0.0651, 0.6703, 474.8877~\text{K}]^{\top}$.
The initial state $x_0$ is uniformly sampled within the range of $[x_s,1.2x_s]$. 
Considering the different magnitudes of the nine state variables of this process, disturbances of different magnitudes are added to the nine states. 
System disturbances $w$ to the original system (\ref{paper1:model:nonlinear}) are generated following Gaussian distributions, and are then made bounded. Specifically, the sequence of disturbances added to each state related to mass fractions is generated following $\mathcal{N}(0,0.5)$, and then made bounded within $[-5, 5]$, while the disturbance added to each state related to temperature is generated following $\mathcal{N}(0,10)$, and is then made bounded within $[-10,10]$.

The state measurements are sampled with a sampling period of $\Delta = 0.001 $ hours. We use~(\ref{paper1: cstr:equation}) to generate datasets for the Koopman model training through open-loop process simulations. The heat inputs $Q_i,i=1,2,3$ are set to random values with added Gaussian noise $\mathcal{N}(0,0.1\times I_3)$ which is bounded by $[-1, 1]\times I_3$. One set of the trajectories of the open-loop control inputs used for data generation is illustrated in Figure~\ref{fig:input}. The levels of the open-loop inputs are generated randomly within a range bounded by $u_{\text{max}}$ and $u_{\text{min}}$, as given in Table~\ref{paper1:table:u}.
\begin{table}[t]
  \renewcommand\arraystretch{1.25}
  \caption{The upper bounds and lower bounds of heating inputs to the three vessels.}\label{paper1:table:u}
  \centering
    \begin{tabular}{ c c c c c c c c c c }
      \toprule
      Bounds of inputs & $Q_1~(\text{kJ/h})$ & $Q_2~(\text{kJ/h})$ & $Q_3~(\text{kJ/h})$ \\
      \midrule
      $u_{\text{max}}$ &
      $3.2 \times 10^6$ & $1.9 \times 10^6$ & $3.2 \times 10^6$ \\
      $u_{\text{min}}$ &
      $2.8 \times 10^6$ & $0.9 \times 10^6$ & $2.8 \times 10^6$ \\
      \bottomrule
    \end{tabular}
\end{table}

While offline data for all the nine process states can be collected for offline modeling, from an online implementation perspective, only the temperatures in the three vessels are measured,  that is, $y=[T_1,T_2,T_3]^{\top}$. 
Therefore, we apply the proposed physics-informed Koopman modeling approach to construct a linear Koopman model to forecast the dynamic behavior of nonlinear system~(\ref{paper1: cstr:equation}). Subsequently, we design a constrained state estimation scheme based on the Koopman model to estimate all the nine system states using the temperature measurements in real-time.

\subsection{Modeling results}

\begin{figure}[!ht]
    \centering
    \includegraphics[width=1\textwidth]{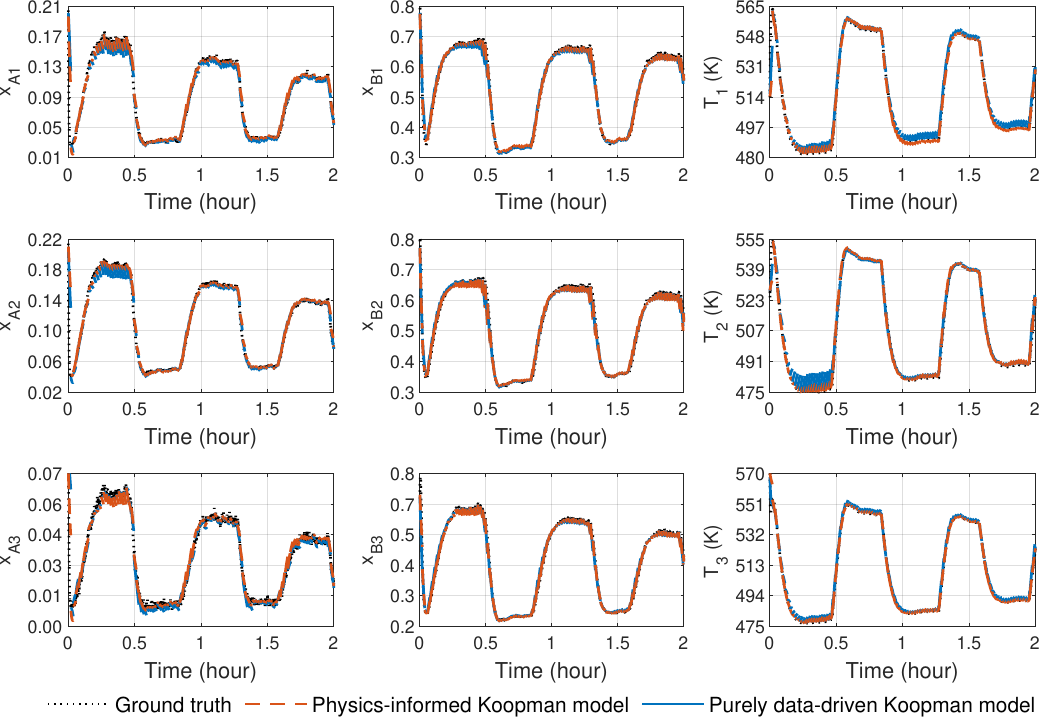}
    \caption{Open-loop Koopman model prediction results.}
    \label{fig:modeling}
\end{figure}

We train the Koopman model using a dataset that contains $2020$ samples, which is able to form $2000$ trajectories of length $20$.  This dataset is then divided into two parts: $80\%$ for training and $20\%$ for validation. An additional dataset containing $2000$ samples is generated as the test dataset. A list of the hyperparameters used in model training is provided in Table~\ref{paper1:table:setting}. The trained Koopman model is used for 20-step ahead open-loop prediction, i.e., $H=20$. The prediction results for the nine states of the process are shown in Figure~\ref{fig:modeling}. The values of all learning rates are determined through mild tuning.

\begin{table}[t]
  \renewcommand\arraystretch{1.25}
  \caption{Hyperparameters for physics-informed Koopman modeling.}\label{paper1:table:setting}\vspace{2mm}
  \centering
    \begin{tabular}{ c c  }
      \toprule
      Parameters & Values  \\
      \midrule
      The size of dataset: $N+H$ & $2020$ \\
      Lifted dimension: $n_l$ & 13 \\
      Activation function & ReLU \\
      Optimizer & Adam \\
      \bottomrule
    \end{tabular}
\end{table}

\begin{figure}[!ht]
    \centering
    \includegraphics[width=1\textwidth]{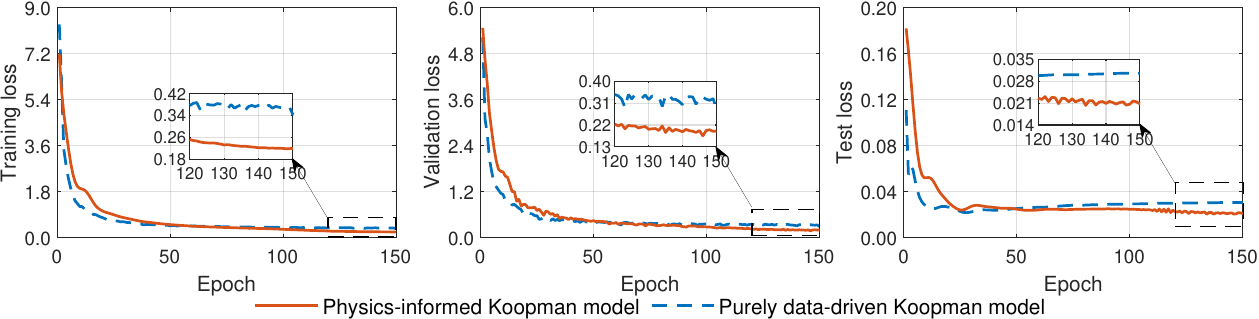}
    \caption{
Trajectories of the training loss, validation loss, and test loss for Koopman model training.}
    \label{paper1:fig:loss}
\end{figure}

Additionally, we compare the proposed physics-informed modeling method with the Koopman modeling method considered in~\cite{baseline1,dlko}. The major difference between our method and the baseline is that our method incorporates physics-informed loss terms into the loss function, whereas the baseline relies only on data. Both models are trained on the same dataset.
We use several different datasets with different initial states and input sequences. For each dataset, we repeat the training process 10 times and compute the average loss values. The results are illustrated in Figure~\ref{paper1:fig:loss}. The comparative modeling result of open-loop Koopman model prediction is shown in Figure~\ref{fig:modeling}. The proposed method provides smaller prediction errors as compared to the baseline.

Through repeated simulations, the training results demonstrate consistent results: 1) The established physics-informed Koopman model does not exhibit overfitting. As depicted in Figure~\ref{paper1:fig:loss}, the test loss for physics-informed model training decreases with epoch, while we see an increasing trend in the trajectory for the baseline method. This may be because the size of the dataset is not sufficiently large for the baseline method;
2) The physics-informed Koopman model achieves smaller minimum values in the training loss, the validation loss, and the test loss. Using mean squared error (MSE) as the criterion, the average test loss for the proposed method is 30\% smaller than that of the purely data-driven baseline. 

\begin{rmk}
The difference in the order of magnitude between the test loss and the other two loss types stems from using different assessment metrics. The train and validation losses are calculated based on the weighted sum of $\mathcal{L}_x$ and $\mathcal{L}_z$, while the test loss is the mean square error between the true and predicted test states, which is the same as $\mathcal{L}_x$. Also, to facilitate the convergence of the trained model, we set an initial weight to each loss term to normalize the order of magnitudes. In this process, we set the weights of $\mathcal{L}_x$ and $\mathcal{L}_z$ as 10. 
\end{rmk}

\begin{rmk}
Based on our simulations, the proposed method consistently provides smaller prediction errors as compared to the baseline. Moreover, the physics-informed Koopman modeling approach can achieve comparable prediction accuracy while requiring 80\% to 90\% fewer samples than a purely data-driven approach.
\end{rmk}

\subsection{State estimation results}

\begin{figure}[ht]
    \centering
    \includegraphics[width=1\textwidth]{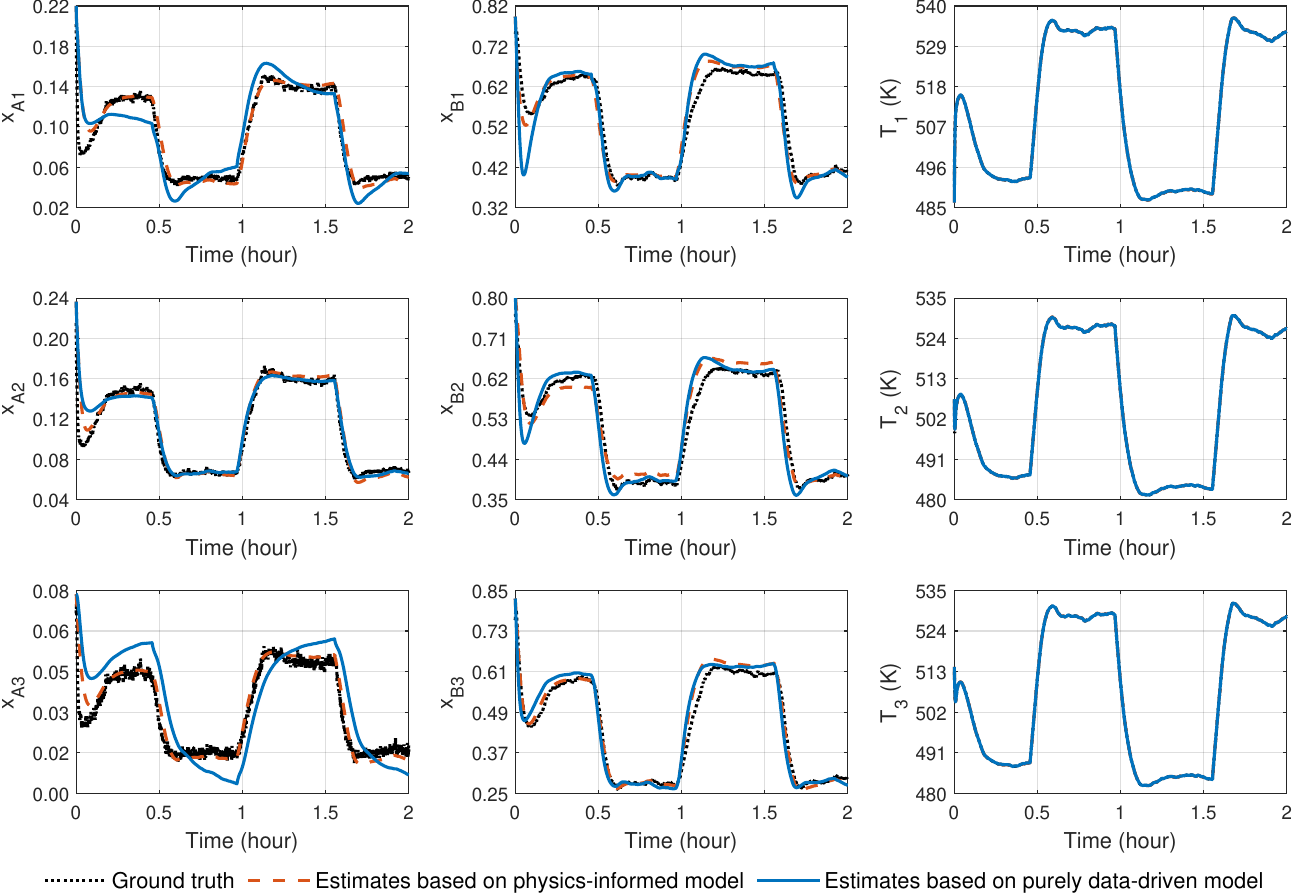}
    \caption{State estimation results based on the physics-informed Koopman model and the purely data-driven Koopman model.}
    \label{fig:mheresult}
\end{figure}

Based on the Koopman model, an MHE scheme with self-tuning weighting matrices is developed.
We generate additional datasets through simulating (\ref{paper1: cstr:equation}) for testing the performance of the MHE scheme. 
In the online implementation, we consider that only the temperatures are online measured, and we estimate all the nine system states. 
The output measurement equation $h$ in~(\ref{paper1:model:nonlinear:2}) is linear, and the matrix $D$ takes the form $D = \bar C C$ where $\bar C_{3\times9}$ has one element of $1$ in each row, corresponding to the temperature measurement in a single vessel, with all other elements being 0.
The initial guess used by the estimator is $\hat{z}_{0|0}=1.2z_0$. The estimation horizon is 40. The estimation results provided by the proposed method are presented in Figure~\ref{fig:mheresult}. The state estimates (dashed lines) accurately capture the actual states (dotted lines).

Additionally, we compare the estimation performance of the proposed method with two other Koopman-based MHE designs. Specifically, we consider three methods: 1)
the proposed self-tuning MHE based on a physics-informed Koopman model, denoted by MHE design 1; 2) MHE based on the physics-informed Koopman model but with constant weighting matrices in the objective function for the MHE, denoted by MHE design 2; 3) an MHE design based on purely data-driven Koopman model and with constant weighting matrices, denoted by MHE design 3.

\begin{table}[ht]
  \renewcommand\arraystretch{1.25}
  \caption{The mean squared error (MSE) for each of the three MHE designs.}\label{paper1:table:mhe}\vspace{2mm}
  \centering
    \begin{tabular}{ c c }
      \toprule
        Method & MSE of the state estimates\\
      \midrule
        MHE design 1 (proposed method) & \bf{0.0412}\\
        MHE design 2 \text{~~~~~~~~~~~~~~~~~~~~~~~~~} & 0.0421\\
        MHE design 3 \text{~~~~~~~~~~~~~~~~~~~~~~~~~} & 0.0972\\
      \bottomrule
    \end{tabular}
\end{table}

The state estimation results given by MHE design 3 are also shown in Figure \ref{fig:mheresult}. While both the proposed method and MHE design 3 can provide estimates that can overall capture the trend of the ground truth, the estimates given by MHE design 1 (the proposed method) are more accurate than MHE design 3. Additional repeated simulations are conducted with different initial states and different input and noise profiles to assess the performance of the three methods. The averaged mean squared errors for the three methods are presented in Table~\ref{paper1:table:mhe}. 
The proposed method achieves the lowest MSE among the three designs, with the MSE of MHE design 1 being reduced by 55\% compared to MHE design 3, and by 2.1\% compared to MHE design 2. The results demonstrate that the use of the physics-informed Koopman model significantly improves state estimation accuracy. Additionally, by dynamically updating the weighting matrices of the Koopman MHE through the proposed method, the accuracy of the estimation is further enhanced.

\section{Data Availability and Reproducibility Statement}
The numerical data used to generate Figures~\ref{fig:input},~\ref{fig:modeling},~\ref{paper1:fig:loss},~\ref{fig:mheresult}, and Table~\ref{paper1:table:mhe} are available in the Supplementary Material. 
The compressed file also contains the simulated data for training, and the corresponding mean and variance used for scaling. 
The results in Figures~\ref{fig:modeling} and Figures~\ref{paper1:fig:loss} are obtained using the provided dataset and the Koopman model established based on the approach proposed in Section~\ref{sec:PIMLmodeling}. Figure~\ref{fig:mheresult} is generated by utilizing the trained Koopman models and the MHE algorithm proposed in Section~\ref{sec:MHE}. We provide the parameters of the Koopman model that are used to generate the open-loop prediction result in Figure~\ref{fig:modeling}. These parameters are then used to build the MHE scheme, and Figure~\ref{fig:mheresult} can be generated. Table~\ref{paper1:table:mhe} is derived by averaging the estimation errors of multiple state estimation simulations conducted based on the established Koopman model.

\section{Conclusion}
\label{sec:conclusion}

In this work, we propose a physics-informed machine learning Koopman modeling approach that leverages both data and available physical information to train the neural network-based lifting functions and Koopman operators. Based on this Koopman model, we formulated a linear self-tuning MHE design to address constrained state estimation of nonlinear systems. The weighting matrices are updated using a pre-trained neural network, which is constructed during the offline Koopman modeling phase. Only convex optimization is required for implementing the MHE-based constrained estimation scheme online, despite the nonlinear dynamics of the considered system.
On a benchmark chemical process, the proposed Koopman modeling and estimation approach is able to show superior performance. The physics-informed modeling method provides enhanced predictive capability and can mitigate overfitting compared to the traditional learning-based Koopman model, especially when dealing with small and noisy datasets. The proposed Koopman-based self-tuning MHE scheme also demonstrates good estimation results. The implementation of the physics-informed Koopman model for MHE and self-tuning of the MHE weighting matrices led to better estimation performance compared to the two baseline Koopman-based MHE methods considered.

\section*{Acknowledgement}
This research is supported by the National Research Foundation, Singapore, and PUB, Singapore’s National Water Agency under its RIE2025 Urban Solutions and Sustainability (USS) (Water) Centre of Excellence (CoE) Programme, awarded to Nanyang Environment \& Water Research Institute (NEWRI), Nanyang Technological University, Singapore (NTU). This research is also supported by the Ministry of Education, Singapore, under its Academic Research Fund Tier 1 (RG63/22 \&RS15/21), and Nanyang Technological University, Singapore (Start-Up Grant). Any opinions, findings and conclusions or recommendations expressed in this material are those of the author(s) and do not reflect the views of the National Research Foundation, Singapore and PUB, Singapore's National Water Agency.


\renewcommand{\refname}{Literature Cited}

\end{document}